# Superconducting stripes induced by ferromagnetic proximity in an oxide heterostructure

Xiangyu Hua[1], Zimeng Zeng[2], Fanbao Meng[1], Hongxu Yao[3], Zongyao Huang[1], Xuanyu Long[2], Zhaohang Li[1], Youfang Wang[3], Zhenyu Wang[1], Tao Wu[1,4], Zhengyu Weng[2], Yihua Wang[3,5], Zheng Liu[2], Ziji Xiang[1*], Xianhui Chen[1,4*].

[1] CAS Key Laboratory of Strongly-coupled Quantum Matter Physics, Department of Physics, University of Science and Technology of China, Hefei 230026, China. [2]Institute for Advanced Study, Tsinghua University, Beijing 10084, China. [3]State Key Laboratory of Surface Physics and Department of Physics, Fudan University, Shanghai 200433, China. [4]Collaborative Innovation Center of Advanced Microstructures, Nanjing University, Nanjing 210093, China. [5]Shanghai Research Center for Quantum Sciences, Shanghai 201315, China.

*Corresponding author. Email: zijixiang@ustc.edu.cn (Z. X.); chenxh@ustc.edu.cn (X. C.).

**The intimate connection between magnetism and superconducting pairing routinely plays a central role in determining the occurrence of unconventional superconducting states. In high-transition-temperature (high-$T_c$) stripe-ordered cuprate superconductors[1-5] and a magnetically ordered iron-based superconductor[6], the coupling between magnetism and superconductivity gives birth to novel phases of matter with modulation of the superconducting pairing in the real space[7-10]. Further exploration of these phases can shed light on the mechanism of unconventional superconductivity. Here we report on the discovery of a peculiar spatially-varying superconducting state residing at the interface between (110)-oriented $KTaO_3$ and ferromagnetic EuO. Electrical transport measurements reveal different $T_c$ and upper critical fields ($H_{c2}$) with current applied along the two orthogonal in-plane directions. Such anisotropy persistently occurs in the low-carrier-density samples that are characterized by strong coupling between Ta $5d$ and Eu $4f$ electrons, whereas in the high-carrier-density samples the coupling is weakened and $T_c$ and $H_{c2}$ becomes isotropic. Complemented by local imaging of diamagnetism and theoretical analysis, our observations imply an unprecedented emergence of superconducting stripes wherein the phase coherence is established ahead of the rest of the interface, arising from a band-filling-dependent ferromagnetic proximity. The realization of such exotic superconducting states provides impetus for the study of novel physics in heterostructures possessing both magnetism and superconductivity.**

Unconventional superconductivity usually emerges in proximity to magnetism. In lanthanum-based copper oxides (the La-214 cuprates), one of the most celebrated families of high-$T_c$ cuprate superconductors, the introduction of holes in the antiferromagnetic insulating parent compound $La_2CuO_4$ can give rise to a unique "stripe phase"[1-4]. In this phase, charge and spin stripes align unidirectionally with correlated spatial modulations; such spontaneous phase segregation has been interpreted as a natural consequence of the moving of itinerant holes in the context of long-range antiferromagnetic order[3,4,11,12]. Most intriguingly, in La-214 cuprates, $La_{1.875}Ba_{0.125}CuO_4$ (LBCO-1/8), the superconductivity shows an unusual "dimension-reduction" behaviour: the in-plane superconducting phase coherence develops at the Berezinskii-Kosterlitz-Thouless (BKT) transition temperature $T_{BKT} \sim 16$ K, whereas the zero resistivity appears at much lower $T$ for current $I \parallel c$, suggesting a $T_c^{3D} \sim 5$ K for the establishment of a three-dimensional (3D) global superconducting phase[5,13].

Such a dimension-reduction phenomenon in a superconducting state is highly unusual, because it challenges the conventional wisdom that $T_c$ is an intrinsic property of a superconductor, which is essentially independent of the direction of applied electric currents (when they are much lower than the critical current). In LBCO-1/8, the emergence of the two-dimensional (2D) superconducting phase with negligible interlayer coupling is closely related to the occurrence of charge and spin stripe orders[1,5,8,14]. Furthermore, it has been pointed out that the interlayer decoupling stems from the intertwinement between the stripe orders and superconductivity, which may lead to a unique pair-density-wave (PDW) order with spatially oscillating superconducting order parameter and finite-momentum pairing[8-10,15]. More recently, a unidirectional PDW order has been revealed inside the superconducting phase in an iron-based high-$T_c$ superconductor $EuRbFe_4As_4$, in which the onset of PDW coincides with the ordering temperature of the Eu magnetic moments[6]. It is widely accepted that the complex interplay between spin orders and superconductivity promises novel physics[7-10,16,17] that may help to understand the enigmatic high-$T_c$ superconductivity[4,10,15,18]. In particular, a system hosting both magnetism and superconductivity, wherein the coupling between them can be switched "on" and "off", is urgently demanded to clarify how such interplay gives rise to unconventional phases of matter.

Here we show that the 2D electron gas (2DEG) formed at the interface between two insulating oxides, i.e., the cubic perovskite $KTaO_3$ (KTO, crystal structure shown in Fig. 1a) and ferromagnetic EuO, can serve as a good candidate for exploring the compromise between magnetism and superconductivity, which is effectively controlled by the interface band filling. Superconductivity and anisotropic normal state resistance have been reported for the 2DEG at the EuO/KTO (111) interface; the latter is supposedly suggestive of an emergent stripe phase[19], yet its impact on the superconducting state still awaits investigations. In this work, we focus on the EuO/KTO (110) interface (Extended Data Fig. 1) grown by means of molecular beam epitaxy. The anisotropy of the KTO (110) surface along the two orthogonal directions, *i.e.*, [001] and [1$\bar{1}$0] (inset of Fig. 1a), is implicit in the crystal structure. Such in-plane anisotropy is clearly manifested in the 2D sheet resistance: as displayed in Fig. 1d, the resistance measured with current *I* applied along [001] is higher than that with *I* along [1$\bar{1}$0].

We prepared EuO/KTO (110) heterostructures with varying sheet carrier density ($n_s$) for the 2DEG at the interface. Control of $n_s$ is achieved by tuning the growth parameters (See Methods for details). Despite the various growth conditions, the EuO overlayers have good single orientation and high surface flatness in our samples (Extended Data Fig. 2). Superconductivity emerges at low temperatures (*T*) in devices with $n_s$ in the range of (6.6-9.6) × $10^{13}$ cm$^{-2}$ and $T_c$ generally increases with increasing $n_s$ (Extended Data Fig. 5). To be noted, the EuO overlayer is ferromagnetically ordered below $T_{curie}$ ~ 73 K (Extended Data Fig. 3a). Therefore, the superconducting interfacial 2DEG is in immediate vicinity of strong ferromagnetism, which represents a key difference from the well-studied LaAlO$_3$/SrTiO$_3$ (LAO/STO) heterostructure.

To eliminate artifacts caused by geometrical effects (see Methods), we fabricated standard Hall-bar devices (Figs. 1b and c; see Methods for technical details) on top of our heterostructure samples. The Hall-bar pattern (Fig. 1b) is designed to simultaneously measure the resistance for the two orthogonal in-plane current directions (i.e., [001] and [1$\bar{1}$0]) on one device. Figures 1e and f display the resistive superconducting transitions measured on the Hall bars patterned onto two samples with $n_s$(3 K) = 9.6 × $10^{13}$ cm$^{-2}$ (high-$n_s$) and 6.8 × $10^{13}$ cm$^{-2}$ (low-$n_s$), respectively. Surprisingly, while $T_c$ is independent of the current direction for the high-$n_s$ 2DEG (Fig. 1e), a notable variation in $T_c$ occurs in the low-$n_s$ 2DEG (Fig. 1f) upon changing the current direction: using the criterion of 50% of the normal state resistance, $T_c$ is 0.62 K for *I* ∥ [001] and 0.44 K for *I* ∥ [1$\bar{1}$0]. This renders differentiation of two phase transitions---the normal state first loses resistance along the [001] direction, and then along the [1$\bar{1}$0] direction at a lower temperature. The existence of two superconducting transitions also manifests in the magnetoresistance (MR) measurements (Extended Data Figs. 6 and 7). The robustness of our observation is further verified by repeating the measurements on multiple Hall-bar devices with different dimensions (Extended Data Fig. 4), as well as devices in the van der Pauw (VDP) geometry (Methods, Extended Data Figs. 5 and 6). Two clearly-separated phase transitions are observed in all samples with carrier density below ~8 × $10^{13}$ cm$^{-2}$ (referred to as low-$n_s$ hereafter), which, however, merge into one transition in samples with higher $n_s$. We note that since the applied currents in all these experiments are well below the critical currents for the superconducting 2DEGs (Extended Data Fig. 8), the pair-breaking effect caused by the current is negligible (Methods).

The upper critical fields ($H_{c2}$) in low-$n_s$ 2DEGs at the EuO/KTO (110) interfaces also display unique features. As a benchmark, in high-$n_s$ devices (Figs. 2a-c), $H_{c2}$ can be well described by Tinkham's 2D model[20] and the current-direction dependence is negligible. In contrast, in low-$n_s$ devices, both the out-of-plane (Fig. 2d) and the in-plane (Figs. 2e and f) $H_{c2}$ depend on the current direction: $H_{c2}$ measured with $I \parallel [001]$ is higher than that with $I \parallel [1\bar{1}0]$ for all field orientations and temperatures, resulting in a broad region in the $H$-$T$ phase diagrams wherein superconductivity is developed in the [001] direction but is absent in the perpendicular direction (orange shaded areas in Figs. 2d-f). This phenomenon further highlights the directional dependence of superconductivity in low-$n_s$ samples. Two additional points are worth mentioning: (i) The Pauli paramagnetic limit ($\mu_0 H_P$ [Tesla] $\approx 1.85$ $T_c$ [Kelvin])[21] is drastically exceeded in both the low-$n_s$ and the high-$n_s$ samples; such a violation, as also reported in other superconducting oxide interfaces/heterostructures[21-24], can be presumably assigned to strong spin-orbit scattering[21,25,26]; (ii) In low-$n_s$ samples, a peculiar curvature change of $H_{c2}(T) \parallel [1\bar{1}0]$ occurs close to $T_c$, which cannot be naturally fitted by the Tinkham's formula (essentially based on the 2D Ginzburg-Landau theory)[20]. Similar behaviours were previously known to occur due to additional complexities, such as multiband effects[27] (see Methods and the fitting curves in Fig. 3f) and unconventional pairing[28]. It is crucial to notice that the $T$ dependence of $H_{c2}$ along [001] can still be well described by Tinkham's formula. We consider that such directional anomaly is intimately connected with the observed directional $T_c$, and a transparent theoretical interpretation is of great interest.

We emphasize that the directional dependence of both $T_c$ and $H_{c2}$ relies strictly on the $n_s$ at the EuO/KTO (110) interface: there appear to be two distinct doping regimes with the presence and absence of such effects, respectively, separated by $n_s \sim 8 \times 10^{13}$ cm$^{-2}$ (Extended Data Figs. 5-7). Such a correspondence with $n_s$ implies that the anisotropic $T_c$ and $H_{c2}$ cannot be fully attributed to the inequivalence of [001] and [1$\bar{1}$0] directions on the KTO (110) surface, or extrinsic factors such as spatially structural/chemical inhomogeneity (Methods). Another potential source of anisotropy stems from the magnetic structure of EuO. Scanning SQUID magnetometry (see Methods for details) shows that the distribution of ferromagnetic domains in the EuO films is random after zero field cooling; the domains are characterized by irregular shapes and typical sizes of 5-20 μm (Fig. 3a); the overall patterns do not change with decreasing $T$ below 700 mK (Extended Data Fig. 10). No stripe-like structure is visible. We further perform scanning SQUID susceptometry, which maps the magnetic response of the heterostructure to a local small alternate current magnetic field[29]. As shown in Fig. 3b, the susceptometry image indicates a spatially homogeneous distribution of magnetic susceptibility at the micrometer level. Thereby, the observed directionality of the interface superconductivity does not arise from peculiar magnetic structures of EuO. More interestingly, the temperature dependence of the susceptibility curves measured at two different spots about 2 mm apart (Extended Data Fig. 10a) both show two successive diamagnetic transitions (Fig. 3c). The transition temperatures are consistent with the zero-resistance temperatures for the two in-plane current directions ([001] and [1$\bar{1}$0], Fig. 1f). This result unambiguously proves that (i) the presence of two superconducting phases with different $T_c$ within the spatial resolution of scanning SQUID (a few microns); (ii) a uniform distribution of such an interfacial superconducting phase coexistence across the sample.

Combining the transport and scanning SQUID data, our experiments strongly suggest the emergence of a novel superconducting state at the EuO/KTO (110) interface. The correspondence between two superconducting transitions in susceptibility and zero-resistance temperatures along the two orthogonal directions imply an interesting scenario: phase-coherent Cooper pairs first form along [001], giving rise to discrete superconducting channels that run across the whole interface, which cause zero resistance only for current paths in this direction. In analogy to LBCO-1/8[5,13] and ZrTe$_3$[30,31], wherein superconducting states with dimensionality lower than the crystal lattice (*i.e.*, the dimension reduction) were reported, here, we propose that (quasi-)1D bundles or "stripes" with $T_c^{stripe}$ higher than that of the 2DEG emerge at the low-$n_s$ interfaces; these "stripes" align along [001] as shown in Figs. 3d and e. In this sense, the $T_c$ detected with $I \parallel$ [001] is $T_c^{stripe}$ at which phase-coherent Cooper pairs form (and are confined) in the stripes, yet the weak coupling between the stripes prevents the establishment of supercurrent in the perpendicular direction (Fig. 3d); upon further cooling, 2D phase coherence develops in between the stripes at $T_c^{2DEG}$, which marks the $T_c$ measured with $I \parallel$ [1$\bar{1}$0]. The stripe-like structures should persist to the lowest temperature (Fig. 3e), as indicated by the fact that $H_{c2}$ at the zero-temperature limit still depends on the current direction (Figs. 2d-f). The two transitions in local susceptibility (Fig. 3c) suggest that the width of the stripes is most likely at the nanoscale (a few hundred nanometers at most).

A key question, however, is in what way the low-density sample is special. Magneto-transport measurements in the normal state provide some insights: in the high-$n_s$ 2DEG the Hall effect is $H$-linear, electron-like and featureless (Fig. 4a), whereas an anomalous Hall effect (AHE) can be resolved in the low-$n_s$ 2DEG at low $T$ (Fig. 4b); the onset of the AHE (~70 K) roughly coincides with the ferromagnetic ordering temperature of the EuO overlayer (Extended Data Fig. 3). Moreover, in-plane MR for low-$n_s$ 2DEGs displays a bow-tie-shaped hysteresis loop at low $H$, which shrinks with increasing $n_s$ and becomes invisible above $n_s \sim 8.6 \times 10^{13}$

cm$^{-2}$ (Fig. 4c) [note that the MR hysteresis is absent for $H \parallel$ [110] (Extended Data Fig. 3d), consistent with the easy-plane anisotropy of the EuO magnetism (Extended Data Fig. 3b)]. In other words, the low-$n_s$ 2DEGs "see" the ferromagnetism of the EuO overlayer, while the high-$n_s$ 2DEGs do not. It is widely perceived that the proximity effect between ferromagnetism (with appreciable exchange field) and superconductivity tend to induce exotic phases of matters, such as finite-momentum pairing state with periodically oscillating order parameter[7,16,32] and odd-frequency superconductivity[17]. The dimension-reduction phenomenon with emergence of superconducting stripes in our low-$n_s$ samples must also be accredited to the strong coupling with overlayer ferromagnetism.

To further look into such coupling, we performed first-principles calculations within the framework of density functional theory (DFT) (see Methods for details) by constructing a (KTO$_3$)$_5$-(EuO)$_{14}$ supercell (Extended Data Fig. 11). Fig. 4e shows the calculated integrated density of states (DOS) from $E_F$ (the charge neutral point) to a given energy $E$ divided by interface area, and the spin polarization defined as the ratio of the spin-majority DOS and the total DOS within a 100 meV energy window centered at $E$. It is evident that the spin polarization rapidly decreases when $E$ moves up. We note that constraint by computational capability and the complexity of the real interface, the integrated DOS should not be directly taken as the experimentally determined $n_s$ (Methods). Nevertheless, the calculation rationalizes that the ferromagnetic proximity effect is strongly filling dependent within a reasonable range of electron density variation.

This result can be understood by noticing that the Eu $4f$-orbitals only overlap with the bottom of the Ta $5d$-bands. The photoemission spectroscopy (PES)[33] on bulk EuO determines that the top of the fully spin-polarized $4f$-orbital bands, as the highest occupied state, is about 3 eV above the top of the O $2p$-orbital bands. KTO has an experimental bulk gap of 3.6 eV between the Ta $5d$ conduction band edge and the O $2p$ valence band edge[34]. Using the O $2p$ level as the energy reference, the Eu $4f$-orbitals are very close to the bottom of the Ta $5d$-bands. This scenario is supported by our band structure modeling. As illustrated in Fig. 4d, the effect of $d$-$f$ hybridization is clearly manifested around the Fermi level, leading to spin splitting. For higher-energy states, the coupling to the $f$-electrons quickly weakens, qualitatively in good agreement with our experimental observations (Figs. 4a-c).

By projecting the orbital weight of the DFT wavefunctions obtained from the supercell, the conduction bands of the KTO/EuO heterostructure can be nicely traced back to the bulk Ta $5d$-bands of KTO. We note that the $d$-bands of bulk KTO have similar dispersions to those in STO; a simple tight-binding modeling for KTO(110) surface is thus available (Extended Data Fig. 12a, Methods). The comparison between the tight-binding model and the DFT-calculated supercell band structure (Extended Data Fig. 12b) reveals that the modulation of the Ta $d$-bands in proximity to Eu $f$-electrons is not only energy (filling) dependent, but also orbital dependent. Specifically, as shown in Fig. 4d and Extended Data Figure 12, $d_{xy}$ bands contribute most to the spin polarization shown in Fig. 4e; $d_{xz}$ bands are pushed away from the low-energy regime; $d_{yz}$ bands remain as the only orbital component with spin degeneracy around the Fermi level, which at the same time shows a quasi-1D dispersion along the [001] direction (Extended Data Fig. 13). We believe that these unique ingredients will serve as a useful foundation for future theoretical explorations.

It would be of great interest to see if the dimension reduction reported herein can be ascribed to an analogous scenario of the case in La-214 cuprates, wherein the intertwinement between superconductivity and charge/spin stripe orders gives rise to decoupled 2D superconducting planes[9,14,18]; here, dimension reduction corresponds to the emergence of 1D structures at the 2D interface due to the (Ta)5$d$-(Eu)4$f$ coupling, providing the first evidence for superconducting stripes occurring in an interfacial superconducting system in proximity to ferromagnetism. Electrons subjected to magnetic exchange fields potentially pair with finite momentum, provoking superconducting states with periodic spatial variation (e.g., the Fulde-Ferrell state[7,16] and PDW state[10]). These oscillating superconducting states naturally cause dimension reduction, since each ordering vector inherently determines an unidirectional modulation of superconducting gap[6,10]; in a 2D plane, 1D modulations can be established, as suggested for EuRbFe$_4$As$_4$[6] and LBCO-1/8 (in which the direction alternation of PDW ordering vectors between adjacent Cu-O layers severely suppresses the interlayer coupling)[9,14]. Future explorations are awaited to reveal whether our observation of anisotropic $T_c$ can be linked to a unidirectional PDW order. Moreover, the occurrence of two distinct superconducting transitions in magnetic susceptibility (Fig. 3c) might also hint at a multicomponent superconductivity[35]. Our discovery thus opens up a promising new pathway for realizing exotic superconducting phases in heterostructure devices by exploiting the magnetic proximity effect.

## Methods

**Growth of EuO/KTO (110) samples and device fabrication.**

EuO thin films were grown on (110)-orientated KTO single crystal using a molecular beam epitaxy system with a base pressure of $4 \times 10^{-10}$ mbar. The typical in-plane dimension of sample is $5 \times 5$ mm$^2$. Before the growing process, the KTO substrates were pre-annealed at 600 °C for 1 hour and then cooled down to the deposition temperature, which is 400 °C for low-$n_s$ samples ($n_s < 8 \times 10^{13}$ cm$^{-2}$ at 3 K) and 450 °C for high-$n_s$ samples ($n_s > 8 \times 10^{13}$ cm$^{-2}$ at 3 K). The deposition rate of Eu was selected as ~0.2 Å/s, which was calibrated by a quartz-crystal monitor. The oxygen pressure during the growth of the samples was kept at $(1.5\text{-}1.8) \times 10^{-9}$ mbar for low-$n_s$ samples and $(1.8\text{-}2.0) \times 10^{-9}$ mbar for high-$n_s$ samples. For the two samples shown in Fig. 1, the growth condition of the high-$n_s$ sample [$n_s$(3 K) = $9.6 \times 10^{13}$ cm$^{-2}$] was chosen as 450 °C and $2.0 \times 10^{-9}$ mbar, whereas for the low-$n_s$ sample [$n_s$(3 K) = $6.8 \times 10^{13}$ cm$^{-2}$] we used 400 °C and $1.7 \times 10^{-9}$ mbar. After growth, the samples were cooled down to room temperature with no oxygen supply. A thin layer of germanium (with a thickness of 3-4 nm) was prepared to protect the sample from further oxidation when exposed to air.

The Hall-bar devices were prepared using standard optical lithography and Argon etching techniques. The etching thickness is ~60 nm which is much larger than the thickness of EuO films (~7 nm).

**Electrical transport measurements**

The transport measurements were performed in a He -3 cryostat (HelioxVT, Oxford Instruments) and a Quantum Design PPMS-9 with a dilution refrigerator insert. In the present work, we used two different measurement configurations: (i) The van der Pauw (VDP) method (data in Fig. 4 and Extended Data Figs. 5-7). This method is utilized to quantitatively analyze the Hall resistance (Figs. 4a and b). We note that a resistance peak appears just above $T_c$ in the VDP measurements (Extended Data Fig. 5), most apparent with the current along the $[1\bar{1}0]$ direction; this peak is likely to reflect the extrinsic geometric effect of the VDP method[36]. (This geometric effect only occurs at the onset of superconductivity, hence it does not affect the identification of anomalous Hall effect in Fig. 4.) To avoid such artifact, we used (ii) the Hall-bar configuration (Figs. 1b and c). We emphasize that the anisotropic superconducting transitions in low-$n_s$ samples (Figs. 1f and 2d-f, Extended Data Figs. 5 and 6) and the absence of such anisotropy in high-$n_s$ samples (Fig. 1e, Extended Data Figs. 5 and 6) are intrinsic properties of the 2DEG at the EuO/KTO (110) interfaces, i.e., these behaviours do not depend on measurement configurations. To further exclude the influence of electric current density on $T_c$, we used small current excitations in all experiments. The currents applied in all VDP measurements are 0.5 µA, two orders of magnitude lower than the critical current $I_c$ ~ 60-100 µA [at $T$ = 0.2 K[21]] (Extended Data Fig. 8a) for the EuO/KTO (110) devices. The current applied in the Hall bar-method experiments is 0.05µA, which is also much smaller than $I_c$ ~ 4-

8 μA [at $T$ = 0.2 K] (Extended Data Figs. 8b and c). The reduction of $T_c$ caused by $J < J_d$ (pair-breaking current density) is given as $\Delta(T_c)/T_c = [T_c(J) - T_c(J=0)]/T_c(J=0) \sim [(1/4)(J/J_d)]^{2/3}$.[37,38] In real superconducting materials, the critical current density $J_c$, which causes non-zero resistivity, is usually much lower than $J_d$ due to the self-field effect[39]. Thereby, we argue that the anisotropic $T_c$ in the low-$n_s$ samples is not caused by anisotropic current-induced pair-breaking effect.

**Two-band fitting model of $H_{c2}$**

In the dirty limit, taking both orbital and Zeeman pair breaking into account, the equation for $H_{c2}$ in a two-band superconductor can be written in the following parametric form[27,40]:

$$\ln t = -[U_1(h) + U_2(h) + \lambda_0/w]/2 + s\{[U_1(h) - U_2(h) - \lambda_-/w]^2/4 + \lambda_{12}\lambda_{21}/w^2\}^{1/2}, \quad (1)$$

$$H_{c2} = 2\Phi_0 k_B T_c th/\hbar D_0, \quad (2)$$

$$U_{1,2}(h) = \operatorname{Re} \psi[1/2 + (i + D_{1,2}/D_0)h] - \psi(1/2), \quad (3)$$

Where $t = T/T_c$, $\psi(x)$ is the digamma function, $D_1$ and $D_2$ are diffusivities in bands 1 and 2, $\lambda_- = \lambda_{11} - \lambda_{22}$, $w = \lambda_{11}\lambda_{22} - \lambda_{12}\lambda_{21}$, $s = sign(w)$, $\lambda_0 = (\lambda_-^2 + 4\lambda_{12}\lambda_{21})^{1/2}$; $\lambda_{11}$ and $\lambda_{22}$ are paring constants in bands 1 and 2; and $\lambda_{12}$ and $\lambda_{21}$ quantify the interband coupling. In Fig. 2f of main text, taking $\eta = D_2/D_1$, $\alpha = D_0/D_1$, $D_0 = \hbar/2m$, our fit yields $\lambda_{11} = 2$, $\lambda_{22} = 1.65$, $\lambda_{12} = 0.92$, $\lambda_{21} = 0.01$, $\alpha = 1$, $\eta = 0.3$.

**Scanning SQUID measurements**

The scanning SQUID measurements on EuO/KTO (110) heterostructures were carried out in a dilution refrigerator with a base temperature of 20 mK. The nano-SQUID probe we used has the similar structure as described in Ref.[41]. The diameter of the pick-up coil is 2.5 μm and that of the field coil is 10 μm. We used the $q$Plus technique for height control[42]. The spatial resolution is limited by both the size of the pickup loop and the height of it from the sample. During the measurements, the nano-SQUID was fixed at 0.4 μm above the sample for collecting the $T$-dependent susceptibility on a single spot (Fig. 3c). For all the susceptometry and magnetometry images, the scanned area is ~ 50 × 80 μm². The effective area of the pickup loop is 3×3 μm², which can be regarded as the spatial resolution of our experiments.

For the scanning SQUID measurements performed in this work, we employed two different modes to probe magnetic properties. Magnetometry ($\Phi$) is a DC measurement of the magnetic flux through the pickup loop as a function of position, showing the intrinsic magnetization of the sample. Susceptometry ($d\Phi/dI_F$) measures the inductive response of the sample through the pickup loop to a modulation alternate current $I_F$ through the field coil. In practice we applied an $I_F = 100$ μA with a frequency of ~1.2 kHz; the magnetometry and susceptometry measurements were performed simultaneously. Because the pickup loop is parallel to the sample surface, it is only sensitive to the local out-of-plane magnetic field. For the susceptometry mapping on the EuO/KTO (110) samples, the paramagnetic response of the ferromagnetic EuO overlayer is larger than the Meissner diamagnetism of the interfacial superconductor, and thus the total susceptibility is positive even below $T_c$ (Fig. 3b).

**First-principles calculations**

For the first-principles calculations, we started from a $(KTO)_5$-$(EuO)_{14}$ supercell (Extended Data Fig. 11) that is manually constructed by cleaving the clean slabs of the individual bulk structure and connecting the KTaO-terminated KTO (110) surface with O-terminated EuO (111) surface. We chose the common in-plane lattice vectors as $a$ = 3.81 Å, $b$ = 5.97 Å and $a \perp b$, which introduces a strain of ~$\pm 5\%$ to the bulk structures. Periodic boundary conditions were employed in all three directions, effectively expanding the heterojunction into a superlattice. The interfacial atomic structure was then relaxed by energy minimization.

The calculations were performed using the VASP code within the framework of density functional theory (DFT)[43]. We employed the projector augmented wave method[44] and the Perdew-Burke-Ernzerhof (PBE) exchange-correlation functional[45]. To properly describe the Eu 4$f$ orbitals, the LDA+U approach was employed with $U - J$ = 5.52 eV[46]. Due to the constraint by computational capability, the simulation supercell enforces the minimal in-plane periodicity, and the slab thickness along the c axis is much smaller than the experimentally grown film. Also, it is understood that the experimental interfacial structure is subject to additional complexities, such as defects and disorders, which also have significant effects on the carrier density. Hence, the integrated DOS should not be directly compared with the experimentally determined $n_s$.

**Tight-binding model for the KTO (110) surface**

Considering a KTO (110) thin film, the lowest quantum well state has the following dispersion [for detailed modeling, please refer to Eq. (S9) in the Supplementary Information of [47]; we take the limit where the lowest quantum well number much smaller compared to the film thickness, and redefine the zero-energy point arbitrarily]:

$$E^{xy}(k_z, k_M) = 2t_2 \cos k_z - 2\sqrt{2t_1^2 + 2t_1^2 \cos(\sqrt{2}k_M)}$$

$$E^{xz/yz}(k_z, k_M) = 2t_1 \cos k_z - 2\sqrt{t_1^2 + t_2^2 + 2t_1 t_2 \cos(\sqrt{2}k_M)} \quad (4)$$

$t_1$ and $t_2$ represent two types of $d$-$d$ hopping processes as schematically shown in Extended Data Fig. 1. The large $t_1$ (-0.68eV) and small $t_2$ (-0.036eV)[48] give rise to anisotropic band dispersion. The simple modeling, which gives the energy bands shown in Extended Data Fig. 12a, should serve as a useful guide for understanding essential physics.

**Exclusion of extrinsic explanations**

For amorphous LaTiO$_3$/SrTiO$_3$ interfaces, the spatial inhomogeneity across the sample can give rise to anisotropy superconducting temperature[49]. We argue that the $T_c$ anisotropy reported in this work is not due to the same origin for the following reasons. First, in ref.[49], the overlayer is amorphous, and the interface is indeed not crystalline; in sharp contrast, the EuO overlayer in our samples has good single crystallinity and the quality of the interface is much better (Extended Data Fig. 2)[21]. The macroscopic homogeneity of our samples has been verified using the both the Hall-bar and the VDP methods (Extended Data Fig. 4). Second, when a four-probe configuration is used for the measurements on the amorphous interfaces, the superconductivity shows no anisotropy but instead exhibits a multiple transition behaviour[49], whereas our observation of anisotropic $T_c$ in standard Hall-bar configuration strongly suggests a totally different origin. Third, the anisotropy superconductivity at the amorphous interfaces occurs randomly[49]. In our low-$n_s$ samples, the anisotropy of $T_c$ is consistently determined by the direction of applied current related to the KTO orientation: $T_c$ was repeatedly confirmed to be higher with $I \parallel [001]$ than $I \parallel [1\bar{1}0]$ in all the low-$n_s$ samples. The robustness of the current-direction-dependent $T_c$ and $H_{c2}$ in the low-$n_s$ samples (Figs. 1f and 2d-f, Extended Data Figs.4-7) and their invariable absence in the high-$n_s$ sample allow us to attribute these unconventional phenomena to the intrinsic coupling effect between the itinerant electrons at the interface and the magnetism of Eu ions. Fourth, the scanning SQUID data (Figs. 3a-c) unambiguously indicate the uniform coexistence of two superconducting phases at a (sub-)micrometer level over the interface, whereas spatially homogenous magnetic signals appear at larger scales. This obviously prefers the intrinsic emergence of unusual superconducting states over macroscopic/mesoscopic inhomogeneity with structural/chemical causes.

In SrTiO$_3$ single crystals, it has been revealed that a large uniaxial strain can induce the formation of 1D dislocations across the sample, which in turn causes a higher $T_c$ along the direction of dislocations[50]. This scenario *cannot* explain the $T_c$ anisotropy in our EuO/KTO heterostructures. The lattice mismatch between EuO (111) and KTO (110) planes at the interface is estimated to be ~5%. Such a mismatch leads to a slight lattice distortion in the EuO overlayer, but no detectable deformation for the KTO layer wherein superconductivity develops[21]. Furthermore, if the mismatch strain induces dislocation at the interface, it would render $T_c$ anisotropic in all samples regardless of the carrier density, contrary to our observations.

**Data availability**

The data that support the findings of this study are available from the corresponding authors upon reasonable request.

**Acknowledgments** We thank Ziqiang Wang, Wei Hu and Jianjun Ying for valuable discussions. This work was supported by the National Key R&D Program of the MOST of China (Grant

No. 2022YFA1602602), the National Natural Science Foundation of China (11888101), Anhui Initiative in Quantum Information Technologies (AHY160000), and the Innovation Program for Quantum Science and Technology (2021ZD0302802). Computational resources for this work were supported by Tsinghua University Initiative Scientific Research Program. The scanning SQUID measurements are supported by National Key R&D Program of China (Grant No. 2021YFA1400100), National Natural Science Foundation of China (Grant No. 11827805 and 12150003) and Shanghai Municipal Science and Technology Major Project (Grant No. 2019SHZDZX01).

**Author contributions** X. H. and X. C. conceived the experiments; X. H., F. M., Z. H. and Z. L. prepared the samples; X. H. and Z. H. performed the transport measurements; Z. Z., X. L., Z. L. and Z. W. performed the first-principles calculation; H. Y., Y. W. and Y. W. performed the scanning SQUID measurements; X. H., Z. W., T. W., Z. W., Y. W., Z. L., Z. X. and X. C. analyzed the data; X. H., Y. W., Z. L., Z. X., and X. C. wrote the manuscript with input from all authors.

**Competing interests** The authors declare no competing interests.

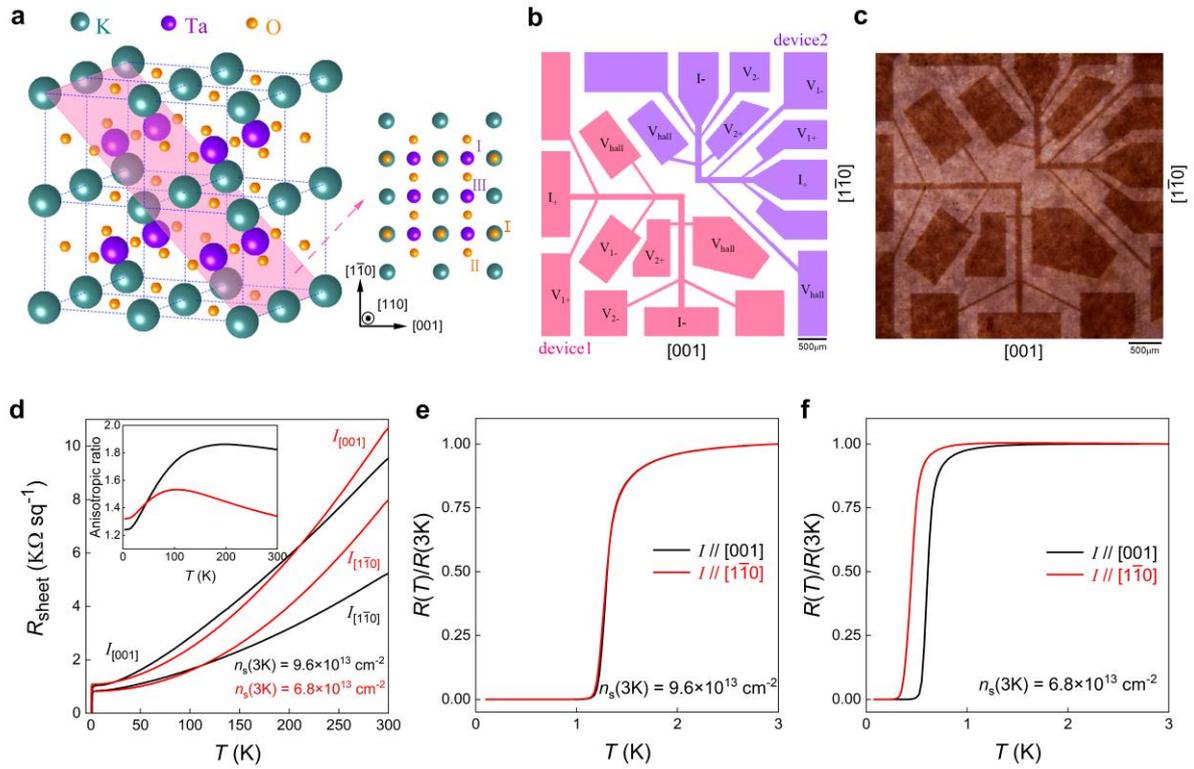

**Fig. 1 | 2DEG formed at the KTO (110) surface and transport measurements on EuO/KTO (110) interface samples. a**, A schematic showing the (110) plane (highlighted by magenta color) in KTO lattice structure; an expanded top view of this plane is shown in the inset, wherein the ions are distributed in three layers (I, II, III). $Ta^{5+}$ and $O^{2-}$ ions are distributed in layers I, III and I, II, respectively. **b**, A schematic diagram of the Hall-bar pattern fabricated on a EuO/KTO (110) heterostructure with a square shape (5mm × 5mm). The pattern contains two devices with different dimensions: device 1 (100μm × 1000μm) and device 2 (100μm × 500μm). **c**, A photograph of the fabricated Hall-bar devices. **d**, The sheet resistance $R_{sheet}$ measured on two Hall bars patterned on samples with different electron density $n_s$ (determined from Hall measurements) plotted against temperature ($T$). Two groups of curves were taken with current applied along the [001] and [1$\bar{1}$0] directions, respectively. Inset: Anisotropic ratio of $R_{sheet}$, defined as $R_{sheet}(I\,\|\,[001])/R_{sheet}(I\,\|\,[1\bar{1}0])$, plotted against $T$ for the two samples. **e, f**, $T$-dependent resistance normalized to the value at 3 K for two samples with (**e**) high-$n_s$ and (**f**) low-$n_s$, respectively, measured using the Hall-bar configuration. Data are shown from 100 mK to 3 K, highlighting the superconducting transitions. The low-$n_s$ sample (**f**) clearly exhibits anisotropic $T_c$ for $I\,\|\,[001]$ and $I\,\|\,[1\bar{1}0]$.

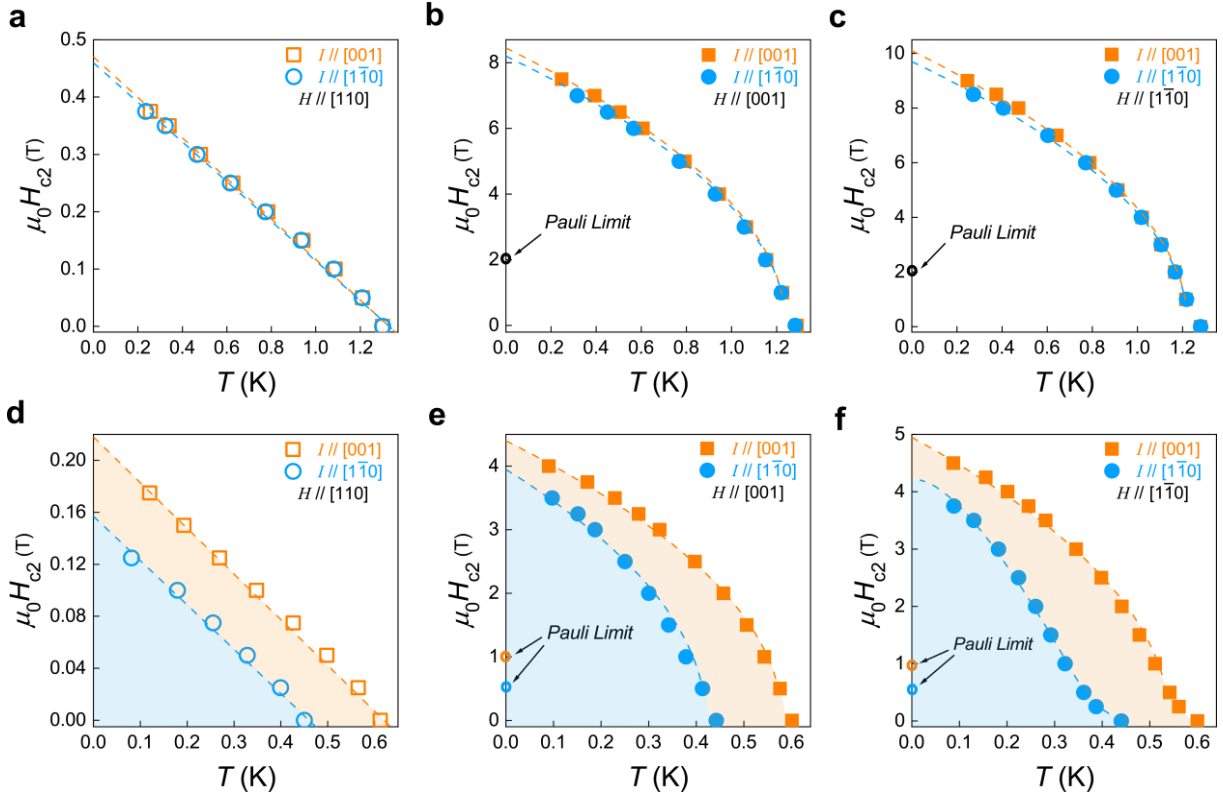

**Fig. 2 | Upper critical fields of the superconducting 2DEGs with high $n_s$ (a-c) and low $n_s$ (d-f) at the EuO/KTO (110) interface**. **a**, **d**, $T$-dependent out-of-plane upper critical fields measured with $I \parallel [001]$ (hollow squares) and $I \parallel [1\bar{1}0]$ (hollow circles). **b-c**, **e-f**, $T$-dependent in-plane upper critical fields measured with (**b**,**e**) $H \parallel [001]$ and (**c**,**f**) $H \parallel [1\bar{1}0]$. Orange and blue symbols denote data for $I \parallel [001]$ and $I \parallel [1\bar{1}0]$, respectively. Upper critical fields are determined by the criterion of 50% normal-state resistance (Extended Data Figs. 9). The zero-temperature Pauli-limit fields $H_P$ estimated from $T_c$ are marked with hollow circles. The dashed lines in **a-e** and the orange dashed line in **f** are fits using 2D Ginzburg-Landau model[20]. The blue dashed line in **f** is a fit of $I \parallel H \parallel [1\bar{1}0]$ data to a two-band model (Methods). Blue and orange shaded areas in **d-f** represent the phase regions in which the superconductivity is established over the whole interface and only in the [001] in-plane direction, respectively. The data were taken in two samples with (**a-c**) $n_s$ (3 K) = $9.6 \times 10^{13}$ cm$^{-2}$ and (**d-f**) $6.8 \times 10^{13}$ cm$^{-2}$. Resistance measurements were performed using the Hall-bar configuration.

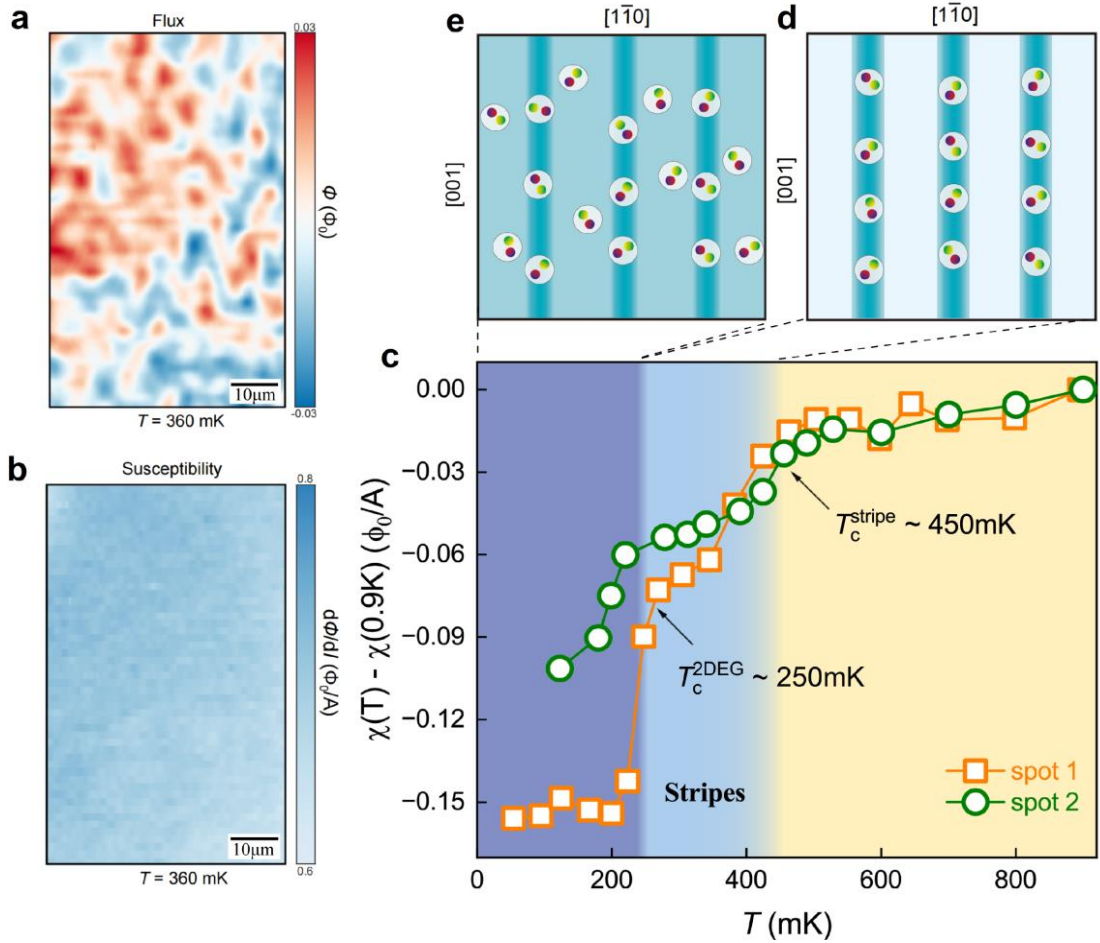

**Fig 3 | Scanning SQUID measurements and schematic diagrams of the superconducting states at the (low-$n_s$) EuO/KTO (110) interface. a**, Magnetometry image measured at 360 mK over the scanned area of 50 × 80 μm² (zone 2 in Extended Data Fig. 10), showing magnetic domains in the EuO overlayer. **b**, Susceptometry image measured at 360 mK reveals a spatially homogeneous magnetic susceptibility in the same zone as in **a**. **c**, $T$ dependence of the susceptibility $\chi$ measured using the scanning SQUID technique at two different positions (spot size 3 × 3 μm², see Extended Data Fig. 10) on a low-$n_s$ sample [$n_s$(3 K) = 6.8 × 10¹³ cm⁻²]. A paramagnetic susceptibility measured at 0.9 K is subtracted from the total signal. Both curves show two diamagnetic transitions whose temperatures are coincident with the $T_c^{stripe}$ and $T_c^{2DEG}$ (defined as the zero-resistance temperatures) in the transport measurements. Shaded areas with different colors denote the temperature regimes divided by the two transitions. The dark blue stripes in **d** and **e** denote the proposed (quasi-)1D superconducting regions formed at the interface, running along the [001] direction. These stripes have a higher $T_c$ than the rest of the 2DEG, *i.e.*, $T_c^{2DEG} < T_c^{stripe}$. White bubbles represent phase-coherent Cooper pairs. **d**, Immediately below $T_c^{stripe}$, superconducting pairing is unidirectional at the interface. Thus, the coherence of Cooper pairs is only established in the [001] direction (along the stripes). **e**, At the lowest temperatures, the entire interface is phase coherent yet anisotropic.

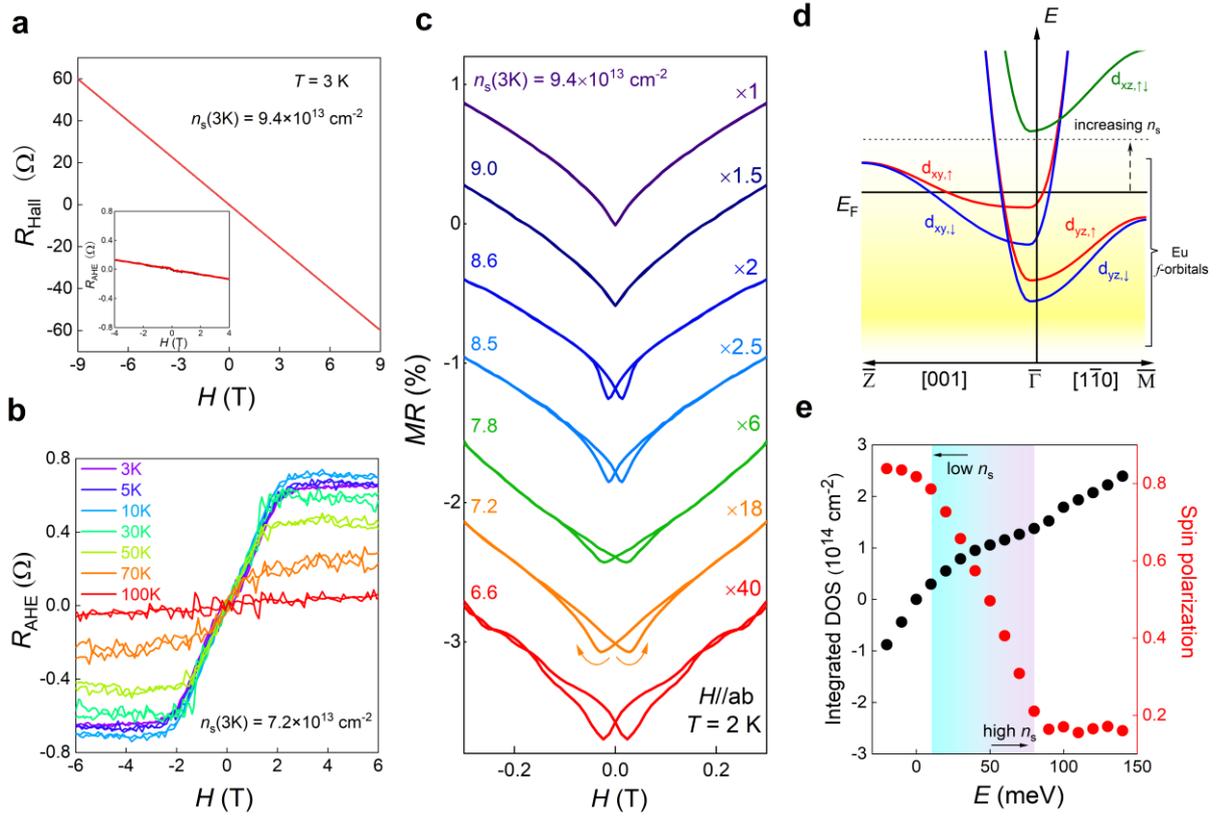

**Fig. 4 | Coupling between the 2DEG and the ferromagnetic EuO overlayer controlled by the band filling. a**, Hall resistance $R_{Hall}$ measured in a high-$n_s$ sample at $T = 3$ K. The slope of $R_{Hall}$ corresponds to $n_s(3\text{ K}) = 9.4 \times 10^{13}$ cm$^{-2}$. Inset shows the residual term of $R_{Hall}$ after subtracting an $H$-linear background; no AHE can be resolved. **b**, The anomalous Hall resistance $R_{AHE}$, obtained in a low-$n_s$ sample with $n_s(3\text{ K}) = 7.2 \times 10^{13}$ cm$^{-2}$ by subtraction of the ordinary $H$-linear term from $R_{Hall}$. $R_{AHE}$ becomes remarkable below ~70 K. At low $T$, $R_{AHE}$ saturates at $H \approx 2.5$ T which is also the saturation field for the magnetization of the EuO overlayer (Extended Data Fig. 3). **c**, MR of 2DEGs with different $n_s$, measured at 2 K with $H$ applied in-plane. Arrows denote the field-sweeping directions. Curves are rescaled and shifted vertically for clearance. **d**, Schematic of band structure of the 2DEG at EuO/KTO (110) interface. Blue and red colors represent two spin components with lifted degeneracy. Green color denotes doubly degenerate bands. The corresponding orbital components are labeled for each band. The yellow shaded area highlights the energy region where the Eu $f$-bands exist (Extended Data Fig. 12). Black dash lines mark out an energy level at which the electron filling (arrow) causes vanishing spin polarization. **e**, Integrated density of states (DOS) and spin polarization derived from the first-principles calculations (Methods) as functions of the Fermi level position $E_F$. The shaded area highlights the energy range in which the spin polarization intensity changes rapidly, consistent with our experiments. Note that the calculated integrated DOS does not quantitatively correspond to the experimental carrier density $n_s$.